\def\useacm{0}

\if\useacm1
\documentclass[sigplan,anonymous,screen,review]{acmart}
\acmSubmissionID{737}
\renewcommand\footnotetextcopyrightpermission[1]{}
\AtBeginDocument{%
\providecommand\BibTeX{{%
\normalfont B\kern-0.5em{\scshape i\kern-0.25em b}\kern-0.8em\TeX}}}

\setcopyright{acmcopyright}
\copyrightyear{2023}
\acmYear{2023}
\acmDOI{XXXXXXX.XXXXXXX}

\acmConference[Eurosys '24]{The European Conference on Computer Systems}{April 2024}{Athens, Greece}
\usepackage{amsthm, amsmath}
\usepackage{bm}
\usepackage{dsfont}
\usepackage{amsfonts}
\usepackage{pifont}
\usepackage{environ}
\usepackage{xcolor, colortbl}
\usepackage{rotating}
\usepackage{engord}
\usepackage{framed}
\usepackage[toc,page]{appendix}
\usepackage{txfonts}
\usepackage{tcolorbox}

\usepackage{algorithm}
\usepackage{algpseudocodex}
\algrenewcommand\alglinenumber[1]{\footnotesize #1:}          
\makeatletter
\renewcommand{\ALG@beginalgorithmic}{\footnotesize}           
\makeatother

\usepackage[per-mode=symbol]{siunitx}

\usepackage{cite}

\usepackage{array}
\usepackage{multirow}
\usepackage{tabularx}
\usepackage{makecell}
\usepackage{diagbox}
\usepackage{booktabs}

\usepackage{url}
\usepackage{breakurl}
\usepackage{xspace}
\usepackage{soul}
\usepackage{caption}

\usepackage{enumitem}
\setenumerate[1]{itemsep=0pt, partopsep=0pt, parsep=\parskip, topsep=5pt}
\setitemize[1]{itemsep=0pt, partopsep=0pt, parsep=\parskip, topsep=5pt}
\setdescription{itemsep=0pt, partopsep=0pt, parsep=\parskip, topsep=5pt}
\setlist[itemize]{align=parleft, left=2pt}
\usepackage{subcaption}
\usepackage{graphicx}

\usepackage{fontawesome5}
\usepackage{tikz}
\usetikzlibrary{calc}
\usetikzlibrary{math}
\usetikzlibrary{patterns}
\usetikzlibrary{arrows.meta}
\usetikzlibrary{decorations}
\usetikzlibrary{decorations.pathmorphing}
\tikzset{
    global scale/.style={
            scale=#1,
            every node/.append style={scale=#1}
        }
}

\makeatletter
\newsavebox{\measure@tikzpicture}
\NewEnviron{scaletikz}[1]{%
    \def\tikz@width{#1}%
    \begin{lrbox}{\measure@tikzpicture}%
        \BODY
    \end{lrbox}%
    \pgfmathparse{#1/\wd\measure@tikzpicture}%
    \BODY
}
\makeatother

\usepackage{lipsum}
\usepackage{blindtext}

\newcommand{\DGC}[1]{{DGC}}

\definecolor{colorpptred}{HTML}{b42611}
\definecolor{colorpptblue}{HTML}{2e6ebf}

\definecolor{cmcolor}{rgb}{0.169, 0.2, 0.51}
\definecolor{codekey}{HTML}{5293c8}
\definecolor{codecomment}{rgb}{0.31,0.49,0.50}
\definecolor{codegray}{rgb}{0.4,0.4,0.4}
\definecolor{codefunction}{rgb}{0,0.13,0.96}
\definecolor{codeconstant}{rgb}{0.92,0.2,0.137}
\definecolor{codeadd}{HTML}{47b53e}
\definecolor{codedel}{HTML}{ae3a2c}

\newcommand{\Me}{\textsc{WeEnv}}

\newcommand{\ione}{(\textit{i})}
\newcommand{\itwo}{(\textit{ii})}

\definecolor[named]{myACMBlue}{cmyk}{1,0.1,0,0.1}
\definecolor[named]{myACMYellow}{cmyk}{0,0.16,1,0}
\definecolor[named]{myACMOrange}{cmyk}{0,0.42,1,0.01}
\definecolor[named]{myACMRed}{cmyk}{0,0.90,0.86,0}
\definecolor[named]{myACMLightBlue}{cmyk}{0.49,0.01,0,0}
\definecolor[named]{myACMGreen}{cmyk}{0.20,0,1,0.19}
\definecolor[named]{myACMPurple}{cmyk}{0.55,1,0,0.15}
\definecolor[named]{myACMDarkBlue}{cmyk}{1,0.58,0,0.21}

\usepackage{listings}
\usepackage{xcolor}
\usepackage{wrapfig}
\usepackage{array}
\usepackage{makecell}

\definecolor{DarkGreen}{RGB}{1,50,32}
\usepackage{wrapfig}

\else
\documentclass[letterpaper,twocolumn,10pt]{article}
\usepackage{usenix2020_09}

\fi

\begin{document}

\title{\Me{}: The Environment for Agentic Reinforcement Learning at WeChat
}
\if\useacm1

    \begin{abstract}
Agentic reinforcement learning (RL) differs from conventional RL in that every task executes inside a complex \textit{environment}, e.g., a virtual machine or
a container.
We find that agentic RL pays a heavy \textit{environment tax}: a large share of the iteration time goes to the environment rather than to learning.
The root cause is the lack of a full-lifecycle solution to environment management.

We present \Me{}, which manages environments across packaging, initialization, and provisioning.
\Me{} packages components as independently published layer groups and composes them at initialization, so that updating a component republishes one small group rather than every artifact containing it.
To speed up environment initialization, \Me{} launches environments instantly and fetches contents on demand.
During task execution, \Me{} provisions CPU and memory elastically, adjusting each environment's quota from its observed usage to fit the varying demands.
\Me{} reduces the initialization by 5.6--14.2$\times$ over E2B, Docker, and AgentENV, cutting its share of the iteration time from up to 53.4\% to 9.1\%.
\Me{} is deployed for agentic RL at WeChat.
\end{abstract}

    
\else
    \author{
{\hypersetup{linkcolor=black}\rm Yang Yu\thanks{Yang Yu, Jing Lei, and Shaoxun Zeng contributed equally to this work.}\quad
Jing Lei\footnotemark[1]\quad
Shaoxun Zeng\textsuperscript{\ensuremath{*}}\thanks{Corresponding author.}\quad
Xinyu Gao\quad Jindi Shi\quad Ci Lei\quad Junjie Zhang}\\
WeChat AI, Tencent
}

\fi

\maketitle

\if\useacm0
    \begin{abstract}
Agentic reinforcement learning (RL) differs from conventional RL in that every task executes inside a complex \textit{environment}, e.g., a virtual machine or
a container.
We find that agentic RL pays a heavy \textit{environment tax}: a large share of the iteration time goes to the environment rather than to learning.
The root cause is the lack of a full-lifecycle solution to environment management.

We present \Me{}, which manages environments across packaging, initialization, and provisioning.
\Me{} packages components as independently published layer groups and composes them at initialization, so that updating a component republishes one small group rather than every artifact containing it.
To speed up environment initialization, \Me{} launches environments instantly and fetches contents on demand.
During task execution, \Me{} provisions CPU and memory elastically, adjusting each environment's quota from its observed usage to fit the varying demands.
\Me{} reduces the initialization by 5.6--14.2$\times$ over E2B, Docker, and AgentENV, cutting its share of the iteration time from up to 53.4\% to 9.1\%.
\Me{} is deployed for agentic RL at WeChat.
\end{abstract}
\fi

\section{Introduction}\label{sec:intro}

Reinforcement learning (RL) has become a key stage in training large language models (LLMs), and its frontier is moving toward long-horizon \textit{agentic} tasks such as software engineering~\cite{swe-bench, swe-smith, swe-rebench, r2e-gym, qwen3-coder} and terminal operation~\cite{terminal-bench}.
RL trains the model in iterations.
Each iteration first performs a rollout, where the current model attempts a batch of tasks and produces trajectories, and then updates the model.

In agentic RL, a task is no longer solved by a single answer.
Driven by a harness, the agent repeatedly calls the LLM for inference and calls tools to act on the task.
Each task hence executes inside an \textit{environment}, a virtual machine or a container whose root filesystem contains files the task execution requires.
The environment is \textit{packaged} offline into an artifact, i.e., a template for virtual machines or an image for containers, and pushed to the registry, a remote service hosting all published artifacts.
At rollout, it is first \textit{initialized} by fetching the artifact and launching an instance, then \textit{provisioned} with resources like CPU and memory while the task executes.

We find that a large share of the agentic RL iteration time is spent on the environment initialization rather than on learning.
Breaking down the iteration time of a production-grade framework, Slime~\cite{slime_github}, we observe that initialization alone accounts for 53.4\% of the iteration time with virtual machines (E2B~\cite{e2b}) and 39.5\% with containers (Docker~\cite{docker}), 2.6$\times$ the task execution itself with E2B and roughly on par with it under Docker, while training takes merely 15.4\%--24.0\%.

The initialization, however, is not the only deficiency.
Current environment management falls short at every stage of the environment lifecycle, and we collectively term these deficiencies the \textit{environment tax}.
First, \textbf{packaging} faces a dilemma.
The environment demands frequent updates, e.g., to evolve the harness~\cite{SWE-agent, adas} and to counter reward hacking~\cite{metr-2025-recent-reward-hacking, anthropic_reward, llm-as-judge-robust}.
\textit{Dynamic assembly} accommodates these updates by installing the evolving components at every initialization, which costs about 20 seconds per environment (\S\ref{moti:packaging}).
\textit{Static bundling} instead bakes them into the artifact, which removes the installation but suffers from a combinatorial explosion of artifacts, where a single update repackages every artifact.
Second, \textbf{initialization} suffers from severe I/O amplification. The environment launches only after its complete artifact is fetched, yet the access of task execution is highly sparse, touching merely 0.81\% of the artifact at the median (\S\ref{moti:init}).
Most of the initialization time is hence wasted on fetching contents that are never used.
Third, resource \textbf{provisioning} allocates a fixed quota, while the peak demands are heavily long-tailed, from about one core at the median to 68 cores at the 99th percentile (\S\ref{moti:provision}), and the provisioning directly affects the latency, inflating the task execution by up to 1.5$\times$ under a limited quota.


We present \Me{}, a full-lifecycle environment solution for agentic RL.
\Me{} consists of three designs.

$\bullet$ \noindent \ul{For packaging, layer composition resolves the dilemma.}
\Me{} packages environment components separately and \textit{composes} them into a complete environment at initialization.
Composition stacks the independently packaged components into the root filesystem of the environment.
It is lightweight, manipulating only metadata, and hence replaces the time-consuming installation of dynamic assembly.
Meanwhile, since components are no longer baked into monolithic artifacts, updating a component republishes one small piece rather than every artifact containing it, avoiding the combinatorial explosion of static bundling.
Such composition is impossible with conventional tooling: only the complete artifact can be published or fetched from the registry, while a layer cannot be located on its own.
\Me{} hence treats layers as first-class deployable units.
Each component is packaged and published as a \textit{layer group} carrying an explicit reference of its own, and an \textit{environment plan} lists the group references and stacks them into the root filesystem at initialization, all on top of the unmodified registry protocol.

$\bullet$ \noindent \ul{For initialization, on-demand fetching removes the I/O amplification.}
\Me{} launches an environment instantly without the complete artifact, and the contents are fetched on demand during task execution.
To support this, \Me{} designs a layer format that separates the layer metadata from the layer contents, so that an environment launches with the metadata alone and any byte range is fetchable on demand.

$\bullet$ \noindent \ul{For provisioning, elastic resource
provisioning replaces the fixed quota.}
Every environment starts from the same small allocation, and \Me{} adjusts its CPU and memory quotas live from the cgroup signals.
Drawing on the characteristics of agentic RL tasks, the scaling follows three principles: scaling up is aggressive while scaling down is conservative, as the demand of a task can surge by $47\times$ within a second (\autoref{fig:F5_task_resource_timeline}), and a wrong reclaim disrupts the task execution and discards the accumulated interaction; scale-ups must not starve the admission of new environments; and a task that cannot be served fails
explicitly rather than degrades silently, so that no corrupted trajectory contaminates the training.

We implement \Me{} and integrate it with Slime.
Across E2B, Docker, and AgentENV~\cite{agentenv_github}, \Me{} reduces the median environment initialization by 5.6--14.2$\times$, cutting its share of the iteration time from up to 53.4\% to 9.1\%.
Layer composition reduces the artifacts to maintain by orders of magnitude.
On-demand fetching launches environments 3.1$\times$ faster and shortens the end-to-end iterations by 1.35$\times$.
Elastic provisioning accelerates the resource-intensive tasks by up to 45$\times$, shortening the iterations they gate by up to 3.7$\times$.
\section{Background}\label{sec:background}
\begin{figure}[!]
    \centering
    \includegraphics[width=\linewidth]{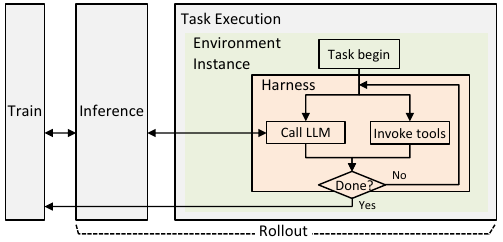}
    \caption{Agentic RL.}
    \label{fig:back-rl-arch}
    \vspace{-0.2cm}
\end{figure}

\noindent \textbf{Agentic RL.}
\autoref{fig:back-rl-arch} shows the architecture of agentic RL~\cite{slime_github, agentgym-rl, agent-lightning}, which iterates between two phases: the rollout phase generates trajectories with the current large language model (LLM), and the train phase updates the model weights with them. 
The rollout involves two parts.
\ione{} The inference part serves the model with an LLM serving engine~\cite{sglang, vllm}.
\itwo{} The task execution part runs each task inside an \textit{environment} instance, driven by the \textit{harness}, i.e., the agent that coordinates the interaction between the model and the environment.
Starting from the task's initial state, the harness repeatedly calls the model through the serving engine and invokes tools inside the environment, e.g., running commands, until the task is judged done.
The interactions form a trajectory, which is sent to the train phase.

\noindent \textbf{Environment in agentic RL.} 
The environment is initialized before task execution, and some tasks~\cite{swe-smith-py, swe-smith,swe-smith-js,swe-smith-cpp,swe-smith-go,swe-smith-rs} require two separate environments.
The first environment hosts the task execution, where the agent solves the task.
After the execution, a second environment is initialized to evaluate the task execution results and get the reward, for those tasks whose reward computation itself requires environment execution.
The evaluation runs in a separate, clean environment, so that rollout-side modifications cannot contaminate the reward computation, guarding against reward hacking~\cite{metr-2025-recent-reward-hacking, anthropic_reward, llm-as-judge-robust}.

Serving as the foundation of task execution, the environment is commonly managed as a separate cluster~\cite{qwen3-coder, kimi-k2, glm-4.5}, for two reasons.
First, since environments are mutated by the agents, concurrent rollouts must be isolated from one another to avoid interference.
Second, the resource profile differs.
Task execution consumes CPU and memory, whereas training and inference are GPU-bound, so pooling environments separately keeps them from fragmenting the GPU capacity.

\noindent \textbf{Environment components.}
The environment is typically instantiated as a virtual machine~\cite{firecracker, e2b} or a container~\cite{docker, swe-bench, SWE-agent, openhands}, which packages the dependencies required for task execution.
It comprises the following components.

\begin{itemize}
    \item The \textit{base} component provides the operating system, language runtimes, shared libraries, and common tools.
    \item The \textit{task definition} component (task for short) comprises a task-specific initial state, e.g., a Git checkout with a faulty patch applied, together with a task specification and an \textit{evaluator} that judges completion and computes the reward.
    The evaluator is updated routinely during RL, as newly discovered reward-hacking behaviors demand increasingly robust evaluation~\cite{metr-2025-recent-reward-hacking, anthropic_reward, llm-as-judge-robust}.
    \item The \textit{harness} component coordinates the interaction between the model and the environment, including model invocation and tool execution; popular harnesses include Claude Code~\cite{claude-code} and Codex~\cite{codex}.
    The harness matters to task performance, as the same model shows different capabilities under different harnesses~\cite{databricks-agent}.
    Moreover, harnesses evolve continuously in their interfaces and tools~\cite{SWE-agent, adas}, requiring the environment to accommodate heterogeneous and evolving harnesses~\cite{agent-lightning, agentgym-rl} for generalization~\cite{qwen3-coder}.
\end{itemize}

\noindent \textbf{Lifecycle of the environment.}
The environment goes through three stages: \textit{packaging, initialization, and provisioning}.
The environment is first packaged offline by the underlying environment tooling into a deployable artifact, i.e., a template for virtual machines such as E2B~\cite{e2b}, or an image for containers such as Docker~\cite{docker}, which is then pushed to a storage service, known as the registry, that hosts all published artifacts.
Notably, the offline-packaged artifact does not necessarily cover the complete environment; in that case, the tooling installs the remaining components during initialization (\S\ref{moti:packaging}).

During the rollout phase in agentic RL, before each task executes, an environment instance is initialized by the same environment tooling, which pulls the artifact from the registry, turns it into the root filesystem, e.g., by unpacking and mounting the artifact for containers, or by loading it as the virtual disk for virtual machines, and launches it.

Multiple environment instances are managed by the \textit{envlet}.
The envlet provisions each instance with resources such as CPU and memory for the duration of task execution.
A node in the environment cluster hosts one or more envlets, allowing flexible resource management.

\section{Motivation}\label{sec:motivation}
In this section, we quantify how environment management impacts agentic RL.

\noindent \textbf{Experiment setups.}
Unless otherwise stated, the experiments in this section use the following setup.
We use Slime~\cite{slime_github}, a popular agentic RL framework, with Megatron~\cite{megatron-lm} as the training engine and SGLang~\cite{sglang} as the inference engine.
Slime by default uses virtual machines (E2B~\cite{e2b}) as the environment; we further extend it to support containers (Docker~\cite{docker}) for a broader study.
We use the SWE-Smith-Python dataset~\cite{swe-smith-py}, Qwen3-8B~\cite{qwen3technicalreport}, and Claude Code~\cite{claude-code} as the harness.
Each experiment runs 256 tasks: 32 unique instances are drawn from the dataset, and each instance is rolled out 8 times with GRPO~\cite{grpo}.

\subsection{The Non-negligible Environment Tax}\label{moti:tax}

\begin{figure}[!]
    \centering
    \includegraphics[width=\linewidth]{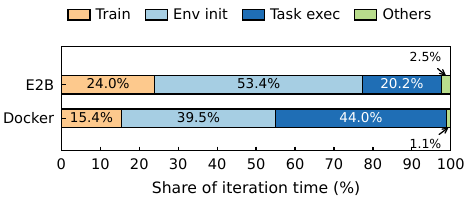}
    \caption{Latency breakdown in agentic RL.}
    \label{fig:F1_train_rollout_other_share}
    \vspace{-0.2cm}
\end{figure}

We break down the iteration time of agentic RL in \autoref{fig:F1_train_rollout_other_share}.
The time of training, rollout, and others is divided by the total latency to obtain the shares.
The rollout is further split into environment initialization and task execution by their ratio summed over all samples.

We find that the rollout, which consists of environment initialization and task execution, dominates the iteration, taking 
73.6\% of the iteration time with E2B and 83.5\% with Docker, while training takes only 24.0\% and 15.4\%, respectively.
Within the rollout, a large share goes to the environment initialization rather than to the task execution.
With E2B, the initialization alone is the single largest component, taking 53.4\% of the iteration time, 2.6$\times$ the task execution itself; with Docker, it accounts for 39.5\%, on par with the task execution (44.0\%).

Such overhead is paid at the initialization stage, but it is partly predetermined at the packaging stage, as analyzed in \S\ref{moti:packaging} and \S\ref{moti:init}.
Beyond packaging and initialization, environment provisioning, which allocates resources such as CPU and memory, decides how fast a task runs, and a limited quota prolongs the task execution (\S\ref{moti:provision}).
These deficiencies of the current environment management collectively limit the pace of agentic RL, and we term them the \textit{environment tax}.

\subsection{Environment Packaging}\label{moti:packaging}
The way the environment is packaged impacts the initialization latency.
As described in \S\ref{sec:background}, components such as the harness evolve continuously, which packaging
must accommodate.

\noindent \textbf{Dynamic assembly.}
One approach is \textit{dynamic assembly}, which defers the accommodation to the initialization stage by excluding the evolving components from the packaged artifact and installing them during the environment initialization.
In Slime's default implementation, which uses virtual machines as the environment, the packaged template includes everything required for task execution except the harness.
After an instance launches from the template, the harness binaries are installed together with the required dependencies, adding significant latency to every environment initialization.
In our measurements, installing the harness adds about 20 seconds to each environment, about 13\% of its initialization time.

Dynamic assembly hence trades initialization latency for
flexibility.
The harness, decoupled from the fixed components, is
free to evolve, but every initialization pays the time-consuming installation.

\noindent \textbf{Static bundling.}
The opposite approach, \textit{static bundling}, packages the complete environment into a single artifact at packaging time.
Since the artifact covers every component, no installation is required at initialization.

Static bundling hence reduces the initialization latency, but suffers from a combinatorial explosion of artifacts.
The Docker backend in our experiments takes this approach.
With $T$ tasks and $H$ harness variants, it requires up to $T \times H$ pre-packaged images, and updating a harness means repackaging and revalidating one image for every task.
Worse, the harness is not the only fast-evolving component; as described in \S\ref{sec:background}, the evaluator is also updated routinely against newly discovered reward hacking.
Each such change touches every affected image, which can take days and lengthens the experimentation cycle of agentic RL.

\definecolor{bluevioletbg}{HTML}{EFF8F1}
\definecolor{bluevioletframe}{HTML}{6F9478}

\newtcolorbox{motibox}{
colback=bluevioletbg,
colframe=bluevioletframe,
boxrule=0.5mm,
arc=1mm,
left=1mm,
right=1mm,
top=0.5mm,
bottom=0.5mm,
before skip=6pt,
after skip=6pt
}

\begin{motibox}
\noindent \textbf{Motivation-1:}
Environment packaging has a direct impact on the subsequent
initialization.
Dynamic assembly sacrifices the initialization latency, while static bundling suffers from a combinatorial explosion of artifacts.
\end{motibox}

\subsection{Environment Initialization}\label{moti:init}
Environment initialization requires a complete artifact in current solutions.
That is, an environment is initialized only after the entire template or image is locally in place.
However, we find that such completeness is unnecessary, since task execution actually accesses only a small part of the artifact.

To quantify the access pattern, we trace all I/O over the root filesystem of each task during execution, which captures the exact contents of the artifact that the task requires.
We report, for each task, the fraction of its artifact accessed during execution, defined as the union of all accessed ranges divided by the artifact size.
Overlapping or adjacent ranges are merged.
For example, $[0, 4\text{ KB})$ and $[2, 8\text{ KB})$ count as one range of 8\,KB.
The metric captures the \textit{sparsity} of the access.
In this experiment we trace on the container runtime based on OverlayFS~\cite{overlayfs}, the same filesystem Docker builds on, which handles writes with copy-up.
Partially writing a file first reads the entire file from the lower layers, so writes also surface as reads, and measuring read I/O alone covers all the artifact bytes that task execution needs.

\begin{figure}[!]
    \centering
    \includegraphics[width=\linewidth]{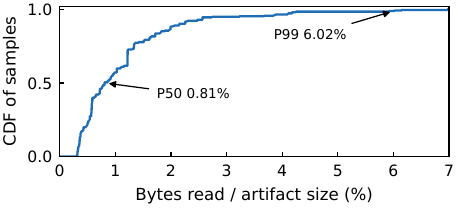}
    \caption{CDF over tasks of the fraction of the artifact accessed during execution.}
    \label{fig:F2_task_image_access_ratio_cdf}
    \vspace{-0.2cm}
\end{figure}

As shown in \autoref{fig:F2_task_image_access_ratio_cdf}, the access is extremely sparse.
The median task reads merely 0.81\% of its artifact, and even the 99th percentile reaches only 6.02\%; no task in our trace touches more than 7\%.
The sparsity arises because a task exercises only what it needs.
For example, a Python library installs many files in the artifact, but executing one task imports only a few of them, and the compilers, headers, and test suites that the artifact carries for other tasks are never opened.

The sparsity means that task execution touches only a small fraction of the environment.
Current solutions, however, fetch the entire artifact before initialization.
The gap between the bytes fetched and the bytes actually accessed manifests as I/O amplification.
For instance, fetching the full image transfers over $100\times$ the bytes that execution ever accesses.

\begin{motibox}
\noindent \textbf{Motivation-2:}
Environment initialization requires a complete artifact in current solutions.
However, task execution accesses only a small fraction of the artifact.
Fetching it in full hence incurs severe I/O amplification.
\end{motibox}

\subsection{Environment Provisioning}\label{moti:provision}

\begin{figure}[!]
    \centering
    \includegraphics[width=\linewidth]{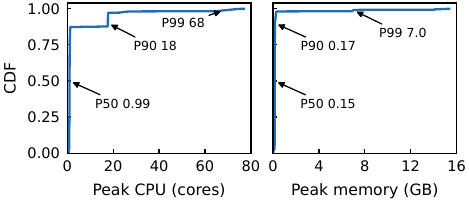}
    \caption{CDFs of the per-task peak CPU and memory usage.}
    \label{fig:F3_task_peak_resource_cdf}
    \vspace{-0.2cm}
\end{figure}

Environment provisioning determines the CPU and memory available to each environment instance during task execution.
To understand the task resource demands, we profile the peak CPU and memory usage of each task, which indicates the resources a task needs to run at its full speed.
\autoref{fig:F3_task_peak_resource_cdf} shows the CDFs over 8k tasks from the SWE-Smith-Python~\cite{swe-smith-py} dataset, and both distributions are heavily long-tailed.
The peak CPU usage is about 1 core at the 50th percentile, but rises to 17.7 cores at the 90th and 68.3 cores at the 99th, a spread of nearly $70\times$.
The peak memory usage is even more skewed.
92\% of the tasks peak below 0.2\,GiB, barely above the median of 0.15\,GiB, yet the 99th percentile jumps to 7.0\,GiB.
A uniform quota hence fits poorly.
One sized for the median starves the tail, while one sized for the tail wastes resources on the majority, e.g., a 7\,GiB quota would over-provision nine out of ten tasks by more than $35\times$.

\begin{figure}[!]
    \centering
    \includegraphics[width=\linewidth]{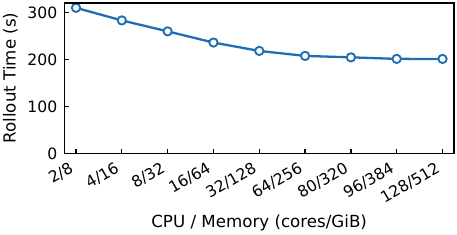}
    \caption{Rollout time of one resource-intensive task under different CPU and memory quotas.}
    \label{fig:F4_quota_exec_time}
    \vspace{-0.2cm}
\end{figure}

We further examine whether provisioning affects the execution time.
We take one resource-intensive task and execute it under different combinations of CPU and memory quotas.
As shown in \autoref{fig:F4_quota_exec_time}, the quota directly
translates into latency.
The rollout takes 310.5\,s under the smallest quota and 201.6\,s under the largest, i.e., a tight quota inflates the rollout by $1.5\times$.
These results show that provisioning more resources directly accelerates task execution.

\begin{motibox}
\noindent \textbf{Motivation-3:}
The resource demands are heavily long-tailed across tasks, while provisioning more resources accelerates task execution.
Hence, provisioning demands differentiated allocation rather than one uniform quota.
\end{motibox}

\section{Our Approach and Challenges}\label{sec:approach_and_challenges}

\noindent \textbf{Layers as first-class deployable units.}
As \S\ref{moti:packaging} shows, environment packaging faces a dilemma.
Dynamic assembly keeps the evolving components decoupled but pays a time-consuming installation at every environment initialization, while static bundling removes the installation at the cost of a combinatorial explosion of artifacts.
Resolving the dilemma requires packaging the evolving components, such as the harness and the evaluator, separately from the rest of the environment, while
\textit{composing} them at the initialization stage without time-consuming installation.

The environment tooling of both virtual machines and containers already embraces a similar notion, the \textit{layer}.
A layer is an independent piece containing several files within the artifact, a natural fit for separating the evolving components from the rest. However, layers serve only as packaging-stage units of caching and deduplication.
At initialization, they must have been bound into a single
monolithic artifact, one template or one image, before an
environment can launch.

To make layers composable at the initialization stage, the
environment tooling must treat the layer, rather than the complete artifact, as the first-class deployable unit.
The fundamental gap is that layers cannot be located on their own.
They exist only as entries in the artifact's manifest, which lists all the layers of the artifact.
Layers thus cannot be indexed apart from the enclosing artifact.
There is hence no way to publish, look up, or fetch a layer independently, let alone compose an environment directly from layers.

\noindent \textbf{On-demand fetching.}
Environment initialization suffers from severe I/O amplification, since task execution accesses only a small fraction of the artifact (\S\ref{moti:init}).
The sparsity suggests a natural opportunity.
Instead of fetching the complete artifact upfront, an environment could launch immediately and fetch artifact contents on demand.

The conventional layer format, however, supports neither side of this opportunity.
Launching an environment before its artifact arrives requires the layer metadata, i.e., the directory tree, the file attributes, and which layer backs each file, to be fetchable separately from the contents; yet a conventional layer is a gzip-compressed tar stream with no central directory, where each file's header sits immediately before its data, so recovering the metadata requires walking the entire layer.
Serving an access on demand further requires reading an arbitrary byte range without fetching the rest of the layer; yet gzip compresses the archive as one sequential stream in which later bytes refer back to earlier ones, so reading any byte implies decompressing everything before it.
Exploiting the sparsity hence demands rethinking the layer format itself.

\noindent \textbf{Elastic resource provisioning during task execution.}
The resource demands are long-tailed across tasks and the execution latency is sensitive to the provisioned resources (\S\ref{moti:provision}), which calls for differentiated allocation.
A naive idea is to predict a task's demands from its beginning: observe the resource usage after the task starts, and adjust for the rest of the execution accordingly.
However, we find that such prediction is impractical, because the demands keep changing throughout the execution.

\begin{figure}[!]
    \centering
    \includegraphics[width=\linewidth]{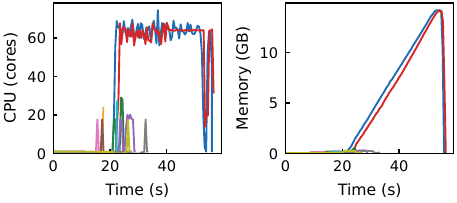}
    \caption{Per-task CPU and memory usage, with tasks aligned at their start.}
    \label{fig:F5_task_resource_timeline}
    \vspace{-0.2cm}
\end{figure}

To examine how the demands evolve, we sample 10 resource-intensive tasks, trace their CPU and memory usage, and align the traces at the task start, as shown in \autoref{fig:F5_task_resource_timeline}.
The usage exhibits no clear pattern: the demand of one task varies drastically over its lifetime, e.g., bursting during compilation and testing while staying idle during model invocation, and the demands differ dramatically across tasks, even between rollouts of the same task.
Moreover, the bursts are abrupt: the CPU demand of one task surges by about $47\times$ within a single second.
The usage observed early in a task hence tells little about the rest, and any one-shot allocation, however informed, misfits some phase of the execution.

\Me{} therefore abandons prediction and provisions \textit{elastically}: the resources of an environment are adjusted dynamically throughout task execution.
\Me{} hence needs to monitor the resource usage of each environment, adjust the quotas live with negligible disruption to the running tasks, and keep the adjustments from starving other environments.
\section{\Me{} Overview}

\begin{figure}[!]
    \centering
    \includegraphics[width=\linewidth]{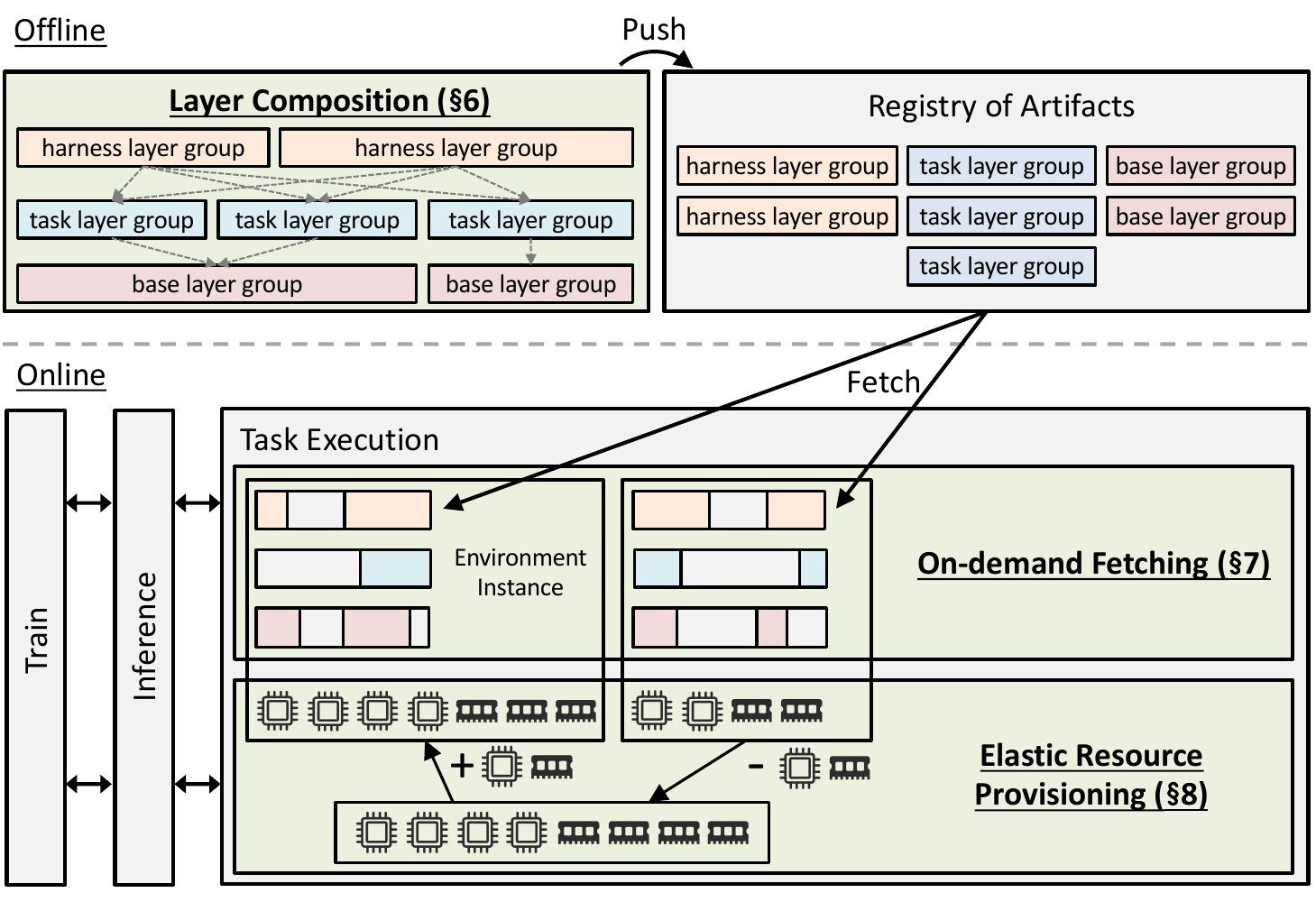}
    \caption{Overview of \Me{}.}
    \label{fig:overview}
    \vspace{-0.2cm}
\end{figure}

We present \Me{}, a full-lifecycle environment solution for agentic RL, from packaging through initialization to resource provisioning.
\Me{} builds on container tooling: each environment is instantiated as a container, whose artifact is an image composed of layers.
As shown in~\autoref{fig:overview}, \Me{} features three key designs.

\noindent \textbf{Layer composition.}
To resolve the packaging dilemma, \Me{} introduces \textit{layer composition}. Environment components are packaged and published separately, each as a \textit{layer group} that may contain several layers, and the groups are composed into a complete environment at the initialization stage.
Updating a component hence republishes one group, rather than repackaging the $T$ artifacts as in static bundling; the updated component takes effect through lightweight composition, rather than the time-consuming installation as in dynamic assembly.

To support the composition, \Me{} gives each layer group an explicit reference and describes an environment with the \textit{environment plan}, an ordered list of such references whose order prescribes how the groups stack into the root filesystem of the environment (\S\ref{sec:layer_index_env_plan}).
The mechanism is built on the unmodified registry protocol, so layer groups are still published to and fetched from standard registries.

Layer composition also offers a new perspective on packaging itself (\S\ref{sec:package_guide}).
The packaging granularity shifts from the complete artifact to layer groups, and the core question becomes identifying the components that change frequently.
Each such component is packaged as a group of its own, so a change republishes only that group rather than the many images containing it.

\noindent \textbf{On-demand fetching.}
Since task execution accesses only a small fraction of the artifact, \Me{} launches environments instantly and fetches artifact contents on demand as task execution accesses them (\S\ref{sec:dataflow}).
The on-demand requests are served from the local and peer caches whenever possible.

Fetching on demand, however, is impossible under the conventional layer format, which interleaves metadata with contents and compresses the layer as one sequential stream.
\Me{} hence designs a new layer format that separates the two (\S\ref{sec:format}): an environment launches with the small metadata alone, and any byte range of a file is fetchable on its own.



\noindent \textbf{Elastic provisioning.}
The resource demands are long-tailed across tasks and vary drastically within a task, so no fixed quota fits.
\Me{} therefore provisions resources \textit{elastically}: every environment starts from a small initial quota, and the envlet monitors its CPU and memory consumption through the cgroup signals, with nothing running inside the environment (\S\ref{sec:monitor_usage}).

From the observed usage, the envlet adjusts the quota live, following three principles (\S\ref{sec:scale_up_and_down}): it scales up aggressively on a single pressure sample while reclaiming only after sustained idleness; it bounds each scale-up and lets an admission preempt the scaled-up quota, so that scale-ups do not starve the environments admitted later; and when an environment still cannot be served, it fails the task explicitly rather than degrading it silently.

\section{Layer Composition}\label{sec:composition}

In this section, we present how layer groups are referenced and composed into an environment through the environment plan (\S\ref{sec:layer_index_env_plan}), and how the composition in turn guides the packaging of the environment (\S\ref{sec:package_guide}).

\subsection{Environment Plan and Composition}\label{sec:layer_index_env_plan}

\begin{figure}[!]
    \centering
    \includegraphics[width=\linewidth]{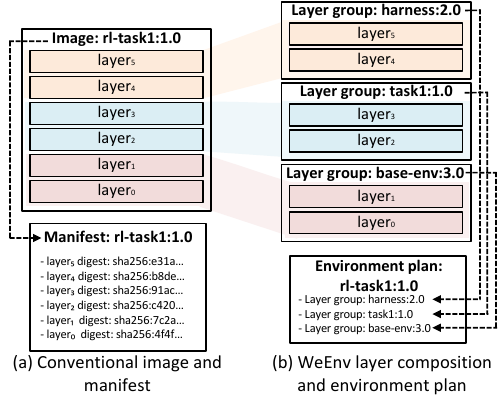}
    \caption{Layer composition and environment plan.}
    \label{fig:layer_composition}
    \vspace{-0.2cm}
\end{figure}

In conventional environment tooling, locating a layer is parasitic on locating an artifact.
Taking the Docker image as an example (\autoref{fig:layer_composition}(a)), the entry point is always an image reference, e.g., \texttt{rl-task1:1.0}.
Layers carry no reference of their own and hence cannot be indexed directly.
Using the image reference, the tooling queries the registry for the image's manifest, which lists its layers, each as a content digest of the layer data.
A digest alone cannot fetch the layer, because the registry accepts only image references.
The only way to fetch a layer is hence to know some image containing it.

\Me{} breaks this dependency by giving each layer group an explicit reference of its own.
As shown in~\autoref{fig:layer_composition}(b), the six layers of \texttt{rl-task1:1.0} split into three independently published groups: \texttt{harness:2.0} (layer$_5$ and layer$_4$), \texttt{task1:1.0} (layer$_3$ and layer$_2$), and \texttt{base-env:3.0} (layer$_1$ and layer$_0$), each versioned and republished on its own.
The same environment is then described by the \textit{environment plan}, which lists the three group references in the stacking order.
Each reference maps to the layer digests of its group through the same mechanism as conventional tooling: the tooling queries the registry for the group's manifest, which lists the digests.
\Me{} also caches the resolved mappings, so a layer group reference is resolved once, rather than repeatedly at every environment initialization, and no knowledge of any enclosing image is required thereafter.

Composition happens at initialization, and it differs from the conventional artifact-based flow in only one step. 
After each group reference is resolved into the ordered layer digests of its group, as described above, the one new step follows, concatenation: the per-group digest lists are joined, from the base group to the topmost, into one flat layer stack, the same shape a conventional manifest prescribes.
The order of the environment plan carries the override semantics: a group placed higher in the plan overrides the groups below it.
The stack is then handed to the mount path, where OverlayFS~\cite{overlayfs} mounts it as a single read-only root filesystem, with each file coming from the topmost group that provides its path; on the very top, OverlayFS places a writable layer private to the environment, which absorbs all writes made by the task.

Composition is hence lightweight: it manipulates only metadata, resolving cached references, concatenating digest lists, and issuing one mount, while no layer content is copied, unpacked, or executed, unlike an installation that runs package managers and writes many files.
The essential change against a conventional artifact is when the stack is formed: the artifact freezes its layer stack at packaging time, whereas the plan assembles it at initialization from independently published groups, so the same groups recompose into different environments without repackaging.

\subsection{Packaging Guided by Composition}\label{sec:package_guide}
Layer composition offers a new perspective on environment packaging.
The packaging granularity shifts from the complete artifact to layer groups, and the core question becomes distinguishing what is subject to change from what is stable.
The guideline is to draw group boundaries along change rates.
Contents that change frequently are packaged as self-contained layer groups, one per independently evolving component, so that each change is published as a new group version on its own and freely composed with the unchanged rest.
Contents that rarely change are packed together and shared across environments.

Following this guideline, \Me{} splits the environment into the following layer groups.
As described in \S\ref{sec:background}, the harness and the evaluator change the most: the harness is continuously upgraded for capability, and the evaluator is routinely patched to counter newly found reward-hacking behaviors.
Each of them is therefore packaged as a layer group of its own.
The relatively fixed contents form the remaining two groups.
The task definition, with the evaluator split into its own group, constitutes the per-task task group, while the operating system, language runtimes, and common tools shared by all tasks constitute the base group.
Under this split, upgrading a harness or patching an evaluator publishes one small group rather than touching any of the $T$ task groups, and the change takes effect in every environment composed afterwards.

Composition can further work together with dynamic assembly.
We observe that many RL tasks~\cite{swe-smith-py, swe-smith,swe-smith-rs,swe-smith-cpp,swe-smith-go,swe-smith-js} share one code repository and differ only in a small mutation, such as checking out a particular commit or applying a faulty patch.
For example, the $39{,}471$ tasks of SWE-Smith-Python span merely $131$ repositories.
Such a mutation is also cheap to perform at the initialization stage.
It completes within a few hundred milliseconds, compared with about 20 seconds for installing a harness (\S\ref{moti:packaging}).
\Me{} therefore packages these tasks at the repository granularity, shared by all tasks on that repository, and applies the per-task mutation dynamically at initialization. This further reduces the number of packaged artifacts from one per task to one per repository, and adds negligible overhead to the initialization stage.
Packaging at the repository granularity also leaves fewer distinct layers across tasks.
Since duplicate content across images is known to be substantial~\cite{vm-image-dedup, dedup-study-msst16}, deduplicating it at the layer level makes the caching more effective~\cite{hands}.
\section{On-demand Fetching}\label{sec:on-demand}

\Me{} launches an environment instantly and fetches the artifact contents on demand as task execution accesses them (\S\ref{sec:dataflow}).
To support efficient on-demand fetching, \Me{} designs a layer format whose metadata and bytes are independently fetchable (\S\ref{sec:format}).

\subsection{Data Flow of On-demand Fetching}\label{sec:dataflow}

To fetch on demand, \Me{} must intercept the file I/O of task execution, and hence mounts the root filesystem into the environment through FUSE~\cite{fuse}.
When task execution accesses a file, the kernel forwards the I/O to the envlet's FUSE handler, which translates it, through the layer metadata, into a request naming the backing layer and the byte range.

The request is served locally whenever possible, through two levels of caches.
The envlet maintains a cache that serves the on-demand requests of all the environments it hosts.
Upon a miss, the envlet turns to the node-level cache; every node in the environment cluster maintains one and shares it with its peers, so a missing chunk can be read from the local node or a peer node.
Only when no node holds the chunk does the envlet fetch it from the remote registry.

Note that only reads enter this path.
\Me{} stacks the layers with OverlayFS~\cite{overlayfs}, where all writes land in the local writable layer.
A partial write to a file backed by the read-only layers triggers a copy-up, which first reads the entire file from the lower layers and hence also surfaces as reads.

In addition, \Me{} runs a best-effort background filling that completes the artifact off the critical path.
It starts once the environment launches, walks the artifact chunk by chunk, and skips the chunks that on-demand fetches have already brought in.
Once a chunk is filled, a later on-demand fetch reads it locally without reaching the remote registry, reducing the stalls in task execution.


\subsection{Layer Format}\label{sec:format}
The designs above pose two requirements on the layer format. First, launching an environment instantly, before its complete artifact arrives, requires the layer metadata to be fetchable separately from the layer contents.
The metadata, i.e., the directory tree, the attributes of each entry, and which layer backs each file, is all that the FUSE mount needs to assemble the root filesystem of the environment.
Second, on-demand fetching requires any byte range of a file to be readable without fetching the rest of the layer.

The conventional layer format meets neither requirement. A layer of Docker, for example, is a gzip-compressed tar stream.
For the layer metadata, it maintains no central directory. Each file's header sits immediately before that file's data, interleaved with the contents, so recovering the layer metadata requires walking the entire layer.
For the contents, gzip compresses the archive as one sequential stream in which later bytes refer back to earlier ones, so reading any byte implies decompressing everything before it.

\Me{} therefore designs a layer format that meets both requirements: the layer metadata is separated from the contents and fetchable on its own, and any byte range of the contents is readable independently.
A layer is converted into this format when its layer group is packaged, and published to the registry as an ordinary blob, so the registry protocol stays unchanged while range reads over the layer become possible.

For the first requirement, the obstacle is that the metadata is scattered throughout the whole layer. \Me{} hence extracts all metadata out of the stream into a standalone table of contents (TOC), placed at the tail of the converted layer.
The TOC records every file's path, attributes, and the byte range of its payload, and is compressed and checksummed on its own.
The envlet obtains the complete metadata of a layer with one size query and one small range read, independent of the file contents; assembling the TOCs of all referenced layers then yields the namespace of the root filesystem, where which layer backs a file follows from the stacking order.

For the second requirement, the obstacle is the compression: gzip permits no seeking.
In the converted format, each file's payload is instead stored contiguously, so any fetched range is directly usable without touching the rest of the layer.

\section{Elastic Resource Provisioning}\label{sec:provision}
\Me{} provisions resources elastically rather than fixing a quota upfront.
Every environment starts from one small initial allocation (2 cores, 8\,GiB).
The envlet then monitors the environment's resource consumption (\S\ref{sec:monitor_usage}) and adjusts its quotas live (\S\ref{sec:scale_up_and_down}), all with negligible disruption to the running task.

\subsection{Monitoring Usage}\label{sec:monitor_usage}
The envlet monitors the CPU and memory consumption of every
environment it hosts.
The envlet manages the resources of each environment through its cgroup, and the same interface serves the monitoring. The envlet periodically reads the cgroup interface files, e.g., \texttt{cpu.stat} and \texttt{memory.stat}, at a 200\,ms interval.
The monitoring is non-intrusive and cheap.
Nothing runs inside the environment, and one round of reads takes only tens of microseconds per environment, consuming less than 0.1\% of one core at the 200-ms interval.

\subsection{Scale Up and Down}\label{sec:scale_up_and_down}
Based on the characteristics of agentic RL tasks, we summarize three scheduling principles that guide scaling decisions.

\ul{First, scaling up is aggressive, while scaling down is conservative.}
The resource demand of a task bursts abruptly, e.g., the CPU demand of one task surges by about $47\times$ within one second (\autoref{fig:F5_task_resource_timeline}).
\Me{} hence scales up aggressively, the moment a single sample crosses a threshold.
Every 200\,ms, the envlet checks each environment against two levels of pressure.
Ordinary pressure doubles the resource quota: for CPU, the utilization reaching 85\% of the quota or 10\% of the scheduling periods being throttled; for memory, the working set reaching 60\% of the quota, or the cgroup reporting reclaim events or pressure.
Severe pressure quadruples the quota at once: for CPU, the
throttling exceeding 80\%; for memory, an OOM emergency.

A wrong scale-down, in contrast, disrupts the task execution, e.g., shrinking memory below the usage kills the environment outright and discards the accumulated turns of model interaction, so scaling down must be conservative.
Scaling down instead requires the usage to stay low for a long time: the CPU quota is halved only after the utilization stays under 30\% for 10 seconds with negligible throttling, and the memory quota only after the working set stays under 20\% of the quota for 60 seconds.

\ul{Second, scale-ups must not starve admissions.}
Some tasks are evaluated in a separate, clean environment
(\S\ref{sec:background}), so one batch admits two waves of
environments: the execution environments upfront, and the evaluation environments as the executions complete.
The two waves overlap on the node, and if the scale-ups of the running environments were allowed to use up the admission capacity, the evaluation environments could not start, making them starve.
\Me{} hence constrains the scale-ups with two mechanisms.
First, a scale-up is bounded.
An environment never grows beyond a per-environment maximum, 64 cores and 32\,GiB by default, and a configured share of the node is reserved for admissions, which no scale-up can consume.
Second, the scaled-up portion can be preempted.
An admission preempts the quotas gained through scale-ups to keep itself from starving, since waiting for the conservative scale-down to return them is too slow.
The envlet immediately reaps the resources that were scaled up for an earlier burst but are no longer used.

\ul{Third, fail explicitly rather than degrade silently.}
A trajectory silently corrupted by starvation is worse than a lost one, as it feeds wrong signals into training.
When an OOM emergency persists with no tier left to raise, the envlet terminates the task and reports the cause, so that the RL framework resamples the task instead of learning from a starved execution.


\section{Evaluation}

\subsection{Setups}

\noindent \textbf{Hardware.}
We run Slime~\cite{slime_github} on four H20 servers, each equipped with eight NVIDIA H20 GPUs (96\,GB HBM each), 384 CPUs, and 2\,TB of memory, where the training engine (Megatron~\cite{megatron-lm}) and the inference engine (SGLang~\cite{sglang}) run colocated.
Two additional servers form the dedicated environment cluster (\S\ref{sec:background}), hosting all environment instances; each is provisioned as a 128-CPU / 512-GiB resource pool with 64 initial environment slots.
The larger-scale experiments (\S\ref{eval:larger_model}) use eight H20 servers instead.

\noindent \textbf{Compared systems.}
We compare \Me{} against three environment backends.

$\bullet$ \texttt{E2B}~\cite{e2b} is Slime's default backend, which instantiates each environment as a virtual machine launched from a pre-built template.

$\bullet$ \texttt{Docker}~\cite{docker} is the container backend we extend Slime to support.

$\bullet$ \texttt{AgentENV}~\cite{agentenv_github} is a recent environment platform for large-scale agentic RL, which instantiates each environment as a Firecracker~\cite{firecracker} virtual machine.
Its root filesystem is served through a ublk userspace block device backed by overlaybd~\cite{overlaybd_github}, a block-level layered image format that converts conventional file-level layers into block-level ext4 layers, enabling block-level on-demand fetching.
Since our servers lack ublk kernel support, the on-demand fetching is disabled in our experiments.
We tune AgentENV for better performance.
The file-level layers are converted into block-level ext4 layers offline and pushed to the registry, so that initialization fetches the converted layers directly and bypasses the conversion during initialization, reducing its initialization latency by 25\%.



\subsection{Overall Performance}\label{sec:eval-overall}

\begin{figure}[!]
    \centering
    \includegraphics[width=\linewidth]{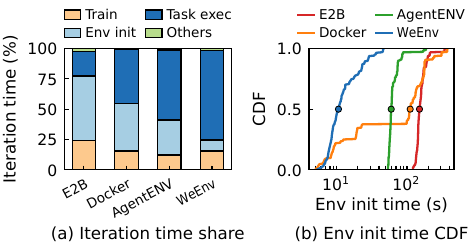}
    \caption{Overall performance under the four environment backends. (a) Iteration time share breakdown. (b) CDF of the per-environment initialization time.}
    \label{fig:F6_e2e_backend_time}
    \vspace{-0.2cm}
\end{figure}

We first evaluate how much \Me{} reduces the environment tax in agentic RL.
We run the same workload on four environment backends, E2B, Docker,
AgentENV, and \Me{}, and report the results in~\autoref{fig:F6_e2e_backend_time}.

\autoref{fig:F6_e2e_backend_time}(a) shows the iteration time
share, following the same breakdown as \autoref{fig:F1_train_rollout_other_share}.
Under the three baselines, environment initialization consumes 28.9\%--53.4\% of the iteration, rivaling or exceeding the task execution itself.
\Me{} cuts this share to 9.1\%, and the task execution rises to 74.0\%, i.e., the iteration finally spends its time on solving the tasks rather than on preparing the environments.

\autoref{fig:F6_e2e_backend_time}(b) explains the reduction from the per-environment view.
The median initialization takes 10.6\,s with \Me{}, against 59.7\,s for AgentENV, 111.0\,s for Docker, and 150.6\,s for E2B, a reduction of 5.6--14.2$\times$.
The reduction follows directly from our designs.
\Me{} launches an environment without waiting for the complete artifact to be fetched, and layer composition eliminates the harness installation that E2B pays at every initialization.
The distributions also differ in shape.
The Docker curve is bimodal: 38\% of the initializations hit a warm container on the node and finish within 30\,s, while the rest pay the full cold creation of 100--300\,s.
\Me{} instead keeps the entire distribution low, as on-demand fetching removes the artifact transfer from the critical path regardless of the cache state.

AgentENV also supports on-demand fetching and caching by design, but differs from \Me{} in two ways.
First, it converts file-level layers into block-level ext4 layers with overlaybd, and the conversion is not cache-friendly: the same layer may map to different blocks across images, so the converted blocks cannot be shared by digest, an issue also observed in file-oriented image services~\cite{flacio}.
\Me{} instead caches at the layer level, where a cached layer is reused wherever its digest matches, regardless of the enclosing image.
Second, AgentENV optimizes solely the initialization, whereas \Me{} spans the full lifecycle, rethinking packaging and resource provisioning as well.

\subsection{Layer Composition}\label{sec:eval-package}
\begin{table}[t]
  \centering
  \caption{Number of packaged artifacts under each packaging strategy with $H$ harnesses.}
  \label{tab:packaging}
  \begin{tabular}{lrrr}
    \toprule
    Dataset & \makecell{Dynamic\\assembly} & \makecell{Static\\bundling} & \Me{} \\
    \midrule
    SWE-Smith-Python     & $39{,}471$ & $39{,}471H$ & $133+H$ \\
    SWE-Smith-C++        & $5{,}123$  & $5{,}123H$  & $71+H$  \\
    SWE-Smith-Go         & $1{,}629$  & $1{,}629H$  & $21+H$  \\
    SWE-Smith-Java       & $6{,}704$  & $6{,}704H$  & $61+H$  \\
    SWE-Smith-JS         & $6{,}073$  & $6{,}073H$  & $36+H$  \\
    SWE-Smith-Rust       & $5{,}311$  & $5{,}311H$  & $41+H$  \\
    SWE-Smith-TS         & $5{,}032$  & $5{,}032H$  & $32+H$  \\
    \bottomrule
  \end{tabular}
  \vspace{-0.2cm}
\end{table}

We evaluate how layer composition reduces the packaging burden.
For each dataset, we count the packaged artifacts, with $H$ denoting the number of harness variants.

As shown in Table~\ref{tab:packaging}, dynamic assembly packages one image per task, and static bundling multiplies this number by $H$, e.g., $39{,}471H$ images for SWE-Smith-Python.
\Me{} instead maintains only $133+H$ artifacts, i.e., around two orders of magnitude fewer.
This reduction is because of layer composition. \Me{} takes the harness and the evaluator out of every artifact, which removes the multiplicative $H$.
\Me{} also packages the remaining tasks at the repository granularity, where tasks sharing one repository reuse a single artifact and the per-task patch is applied at initialization.
In addition to these repository-level task artifacts, \Me{} packages the shared base tools as one artifact, the evaluator as one dataset-specific artifact, and one artifact per harness, yielding $B+2+H$ artifacts for $B$ repository-level base images and $H$ harnesses.

Beyond the counts, the strategies differ in the cost of change.
Updating a harness or patching an evaluator under static bundling touches every affected image, up to the full $T$ per dataset; under \Me{}, it republishes exactly one layer group, and the change takes effect in every environment composed afterwards.
Dynamic assembly avoids the repackaging but pays the installation at every initialization, as analyzed in \S\ref{moti:packaging}.
\Me{} also works together with dynamic assembly, as the per-task patch is applied at initialization.
The overhead is negligible.
The patching completes in 0.43\,s at the median, measured over $1{,}024$ initializations, compared with 20 seconds for installing a harness.
\Me{} hence keeps the flexibility of dynamic assembly and the launch speed of static bundling, while maintaining the fewest artifacts.

\subsection{On-demand Fetching}\label{sec:eval-ondemand-fetch}

\begin{figure}[!]
    \centering
    \includegraphics[width=\linewidth]{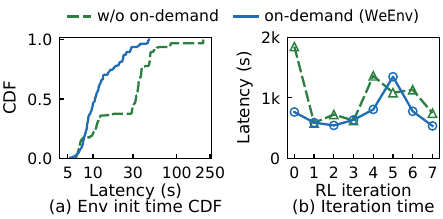}
    \caption{Ablation of on-demand fetching. (a) CDF of the per-environment initialization latency. (b) End-to-end time of each RL iteration.}
    \label{fig:F7_env_init_fetch_mode}
    \vspace{-0.2cm}
\end{figure}

We ablate on-demand fetching by rerunning the same workload with it disabled, so that every initialization downloads its complete artifact before launching (\textit{w/o on-demand}); \Me{} instead launches the environment instantly and fetches layer contents on demand upon first access (\S\ref{sec:dataflow}).

\autoref{fig:F7_env_init_fetch_mode}(a) shows the CDF of the per-environment initialization latency.
With on-demand fetching, the median initialization takes 10.6\,s, 3.1$\times$ faster than the 32.5\,s without it, and the improvement grows toward the tail: the P95 drops from 83.5\,s to 35.1\,s, and the worst case from 208\,s, when the layers must come from the remote registry, to 47\,s. On-demand fetching launches every environment after transferring only the chunks it immediately touches, so the
initialization latency no longer depends on the artifact size.

\autoref{fig:F7_env_init_fetch_mode}(b) shows the end-to-end time of each RL iteration. The gap is largest in the first iteration, where all environments start cold: downloading the full layers for every environment stretches the iteration to 1{,}845\,s, while on-demand fetching completes it in 763\,s, a $2.4\times$ reduction.
Later iterations partially reuse locally cached layers and the gap narrows, yet all iterations still total 8{,}092\,s without on-demand fetching versus 5{,}988\,s with it, a $1.35\times$ speedup.

\subsection{Elastic Resource Provisioning}\label{sec:eval-provisioning}

\begin{figure}[!]
    \centering
    \includegraphics[width=\linewidth]{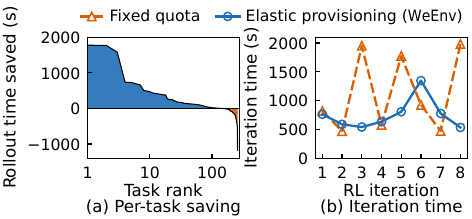}
    \caption{Elastic provisioning versus a static quota. (a) Per-task execution time saved by elastic provisioning, ranked by the savings. (b) End-to-end latency of each iteration.}
    \label{fig:F8_provisioning_ablation_first32_rg32}
    \vspace{-0.2cm}
\end{figure}

We ablate elastic provisioning by rerunning the same workload with it disabled, so that every environment keeps the initial quota of 2 cores and 8\,GiB throughout.
With elastic provisioning, environments start from the same initial quota and \Me{} adjusts it from the observed usage (\S\ref{sec:scale_up_and_down}).

\autoref{fig:F8_provisioning_ablation_first32_rg32}(a) ranks all paired task executions by the rollout time that elastic provisioning saves.
The savings concentrate on the long tail identified in \S\ref{moti:provision}.
61 executions save more than 100\,s each.
When the agent issues resource-hungry commands inside the environment, the 2-core quota stretches them from seconds to minutes, and the slowest two rollouts escalate to the 1{,}800-s agent timeout; with elastic provisioning the same tasks finish in about 40\,s, a $45\times$ reduction.
A few executions at the right end run longer under elastic provisioning.
Their extra time is spent in the solving phase, where the agent takes a different, longer trajectory, a stochastic effect of agent sampling rather than of the resource allocation.

\autoref{fig:F8_provisioning_ablation_first32_rg32}(b) shows how the per-task savings translate into iteration latency.
An iteration waits for its slowest rollouts, so the stalled rollouts in (a) directly gate the iterations that contain them: iterations 3, 5, and 8 take 1{,}949\,s, 1{,}773\,s, and 1{,}981\,s under the fixed quota, and drop to 542\,s, 806\,s, and 534\,s with elastic provisioning, up to $3.7\times$.
Summed over all iterations, elastic provisioning lowers the total rollout latency from 8{,}968\,s to 5{,}988\,s, a $1.50\times$ speedup.

\begin{figure}[!]
    \centering
    \includegraphics[width=\linewidth]{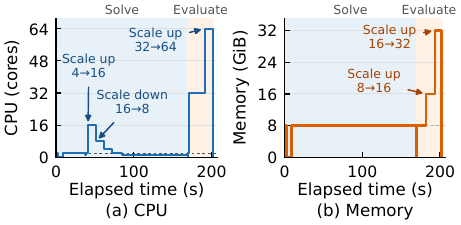}
    \caption{Quota timeline of one resource-intensive task under elastic provisioning. (a) CPU. (b) Memory. Dashed lines mark the initial quota (2 cores and 8\,GiB).}
    \label{fig:F9_first32_elastic_scaling_example}
    \vspace{-0.2cm}
\end{figure}

\autoref{fig:F9_first32_elastic_scaling_example} shows a concrete example of elastic provisioning, tracing the CPU and memory quotas of one resource-intensive task over its lifetime.
The task goes through two phases: solving the task, and running the evaluator to get the reward.

The CPU quota (\autoref{fig:F9_first32_elastic_scaling_example}(a)) starts from the initial 2 cores.
During solving, a burst raises the quota from 4 to 16 cores at once; as the burst subsides, the conservative scale-down halves the quota step by step back to 2 cores, so the released cores serve other environments on the node for most of the solving phase.
The evaluation phase bursts again, and the quota climbs through 32 to the per-environment maximum of 64 cores.
The memory quota (\autoref{fig:F9_first32_elastic_scaling_example}(b)) tells a different story: it stays at the initial 8\,GiB through the entire solving phase, as the phase is dominated by model invocations that consume little memory, and doubles twice to 32\,GiB only when the evaluator demands it.
When the task completes, the environment is torn down and the quotas drop to zero at once, without waiting for the conservative scale-down.
The example confirms the principles in \S\ref{sec:scale_up_and_down}.
Scale-ups react within a single sample, so neither burst stalls the task; scale-downs retreat conservatively, so no adjustment disrupts the execution; and the same environment holds widely different quotas at different moments, which no fixed allocation could match.

\subsection{Sensitivity Analysis}

\subsubsection{Harness}

\begin{figure}[!]
    \centering
    \includegraphics[width=\linewidth]{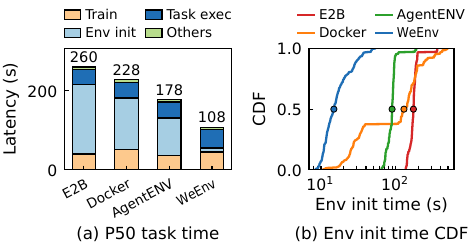}
    \caption{Overall performance with the Codex harness. (a) Median per-task time breakdown. (b) CDF of the environment initialization time.}
    \label{fig:F10_codex_e2e_backend_time}
    \vspace{-0.2cm}
\end{figure}

In this experiment, we change the harness from Claude Code~\cite{claude-code} to Codex~\cite{codex} to evaluate whether \Me{}'s benefits hold under a different harness.
We rerun the overall performance comparison in \S\ref{sec:eval-overall} with the Codex harness.

\autoref{fig:F10_codex_e2e_backend_time}(a) shows the median per-task time.
The conclusion stays consistent with Claude Code.
\Me{} completes the median task in 108\,s, 2.4$\times$ faster than E2B (260\,s), 2.1$\times$ faster than Docker (228\,s), and 1.6$\times$ faster than AgentENV (178\,s).
\autoref{fig:F10_codex_e2e_backend_time}(b) shows the CDF of the per-environment initialization time.
The median initialization takes 13.7\,s with \Me{}, against 87.3\,s for AgentENV, 126.8\,s for Docker, and 160.2\,s for E2B, a reduction of 6.4--11.7$\times$.
These results show that \Me{}'s benefits are not tied to a specific harness.

\subsubsection{Dataset}

\begin{figure}[!]
    \centering
    \includegraphics[width=\linewidth]{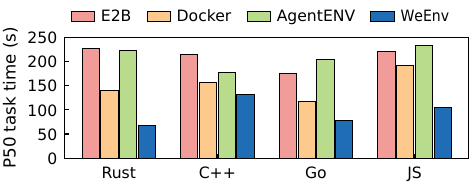}
    \caption{Median per-task time across four SWE-Smith language datasets under the four environment backends.}
    \label{fig:F12_multidataset_e2e_backend_time}
    \vspace{-0.2cm}
\end{figure}

In this experiment, we change the dataset from SWE-Smith-Python to four other language datasets of SWE-Smith~\cite{swe-smith} (Rust~\cite{swe-smith-rs}, C++~\cite{swe-smith-cpp}, Go~\cite{swe-smith-go}, and JavaScript~\cite{swe-smith-js}), to evaluate whether \Me{}'s benefits hold across datasets.
\autoref{fig:F12_multidataset_e2e_backend_time} reports the median end-to-end task time.

\Me{} completes the median task fastest: 66.9\,s on Rust, 131.5\,s on C++, 78.9\,s on Go, and 105.3\,s on JavaScript: 1.2--2.1$\times$ faster than Docker, 1.6--3.4$\times$ faster than E2B, and 1.3--3.3$\times$ faster than AgentENV.
The advantage varies with the dataset in a predictable way.
\Me{}'s saving comes chiefly from the environment side, which is roughly constant per task, so datasets whose tasks execute quickly see the largest relative gains: on Rust and Go the median task drops to 66.9\,s and 78.9\,s, up to $3.4\times$ faster than the baselines.
On C++, whose compilation-heavy execution dominates the task time, the gap against the best baseline narrows to $1.2\times$, yet \Me{} still leads.
These results show that \Me{}'s benefits are not tied to a specific language or dataset.

\subsubsection{Scaling to Larger Models}\label{eval:larger_model}

\begin{figure}[!]
    \centering
    \includegraphics[width=\linewidth]{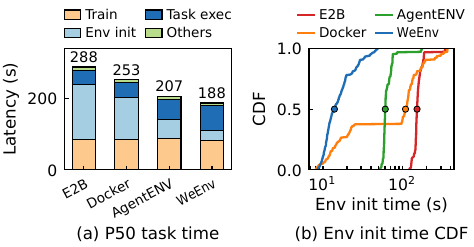}
    \caption{Overall performance with the larger Qwen3-14B model on the eight-server cluster. (a) Median per-task time breakdown. (b) CDF of the environment initialization time.}
    \label{fig:F11_qwen3_14b_e2e_backend_time}
    \vspace{-0.2cm}
\end{figure}

In this experiment, we scale the trained model from Qwen3-8B to Qwen3-14B and move the workload to the eight-server cluster.
We rerun the overall performance comparison in \S\ref{sec:eval-overall} and report the results in~\autoref{fig:F11_qwen3_14b_e2e_backend_time}.

\autoref{fig:F11_qwen3_14b_e2e_backend_time}(a) shows the median per-task time under the four backends.
\Me{} completes the median task in 188\,s, 1.5$\times$ faster than E2B (288\,s), 1.3$\times$ faster than Docker (253\,s), and 1.1$\times$ faster than AgentENV (207\,s).
\autoref{fig:F11_qwen3_14b_e2e_backend_time}(b) confirms this from the CDF of the per-environment initialization time.
The median initialization takes 13.9\,s with \Me{}, against 60.4\,s for AgentENV, 108.2\,s for Docker, and 150.7\,s for E2B, a reduction of 4.3--10.8$\times$.
These results show that, as the model scales, the environment-side savings stay constant in absolute terms, while the initialization speedup holds.

\section{Related Work}
\label{sec:related}

\noindent \textbf{Systems for agentic RL.}
Agentic RL extends the RL loop of LLM post-training from single-turn answers to long-horizon tasks, where a harness drives the model through repeated inference and tool calls inside an environment.
Recent frameworks target this setting, e.g., Slime~\cite{slime_github}, AgentGym-RL~\cite{agentgym-rl}, and Agent Lightning~\cite{agent-lightning}.
They decouple the agent program from the trainer so that an existing harness can be trained without modification.
A parallel line of work optimizes RL from the scheduling side: RLinf~\cite{rlinf} lowers the RL workflow into fine-grained flows that can be placed collocated or disaggregated, RollPacker~\cite{rollpacker} consolidates long-tail trajectories into dedicated batches to keep GPUs from idling, and RollArt~\cite{rollart} disaggregates rollout, reward, and training onto matched hardware.
All of them treat the environment as a black box that is available on demand.
\Me{} instead optimizes the environment across its full lifecycle.

\noindent \textbf{Accelerating environment startup.}
Lazy image distribution is the closest line of work.
Slacker~\cite{slacker} observed that only a small fraction of the image is read, which led to on-demand pulling at the file or block level in eStargz~\cite{estargz}, Nydus~\cite{nydus}, DADI~\cite{dadi}, SOCI~\cite{soci}, and FlacIO~\cite{flacio}, making a container usable before the complete image arrives.
Prefetching guided by predicted working sets~\cite{starlight} further hides the fetch latency when the access pattern is predictable, and SOCK~\cite{sock} streamlines the isolation primitives on the launch path.
\Me{} shares the goal and further exploits the characteristics of agentic RL to reduce what must be published, fetched, and cached.
It packages the fast-evolving components as independently published layer groups and the tasks at the repository granularity, which cuts the artifacts to maintain and leaves fewer distinct layers, making the caching more effective.

Snapshot and fork-based startup forms a second line, mainly for virtual machines and serverless functions. Firecracker~\cite{firecracker} provides the microVM substrate, and FaaSnap~\cite{faasnap}, REAP~\cite{reap}, and PASS~\cite{pass} reduce restore latency by prefetching or pinning the pages that the restored instance actually touches. These techniques target a snapshot taken from a known, unchanged root filesystem. In agentic RL, each task arrives with its own initial state, so \Me{} instead accelerates the path by which the artifact reaches the node.
\section{Conclusion}
This paper reveals that agentic RL pays a heavy environment tax.
We present \Me{}, which manages the entire lifecycle of the environment, letting the iteration spend its time on learning rather than on preparing environments.
\Me{} is deployed for agentic RL at WeChat.

\if\useacm1
    \bibliographystyle{ACM-Reference-Format}
\else
    \bibliographystyle{plain}
\fi
\bibliography{refs.bib}


\end{document}